\documentclass[runningheads]{llncs}
\usepackage{graphicx} 

\begin{document}
\title{Mobile user experience from the lens of project-based learning} 
\author{Maria Spichkova\inst{1}\orcidID{0000-0001-6882-1444}}
\authorrunning{Maria Spichkova} 
\institute{RMIT University, Melbourne, Australia\\
\email{maria.spichkova@rmit.edu.au}\\  
}
\maketitle              
\begin{abstract}
This paper presents an overview of mobile application projects conducted at the RMIT University as a part of the Learning and Teaching activities within Bachelor and Master programs, in collaboration with industrial partners. 
We discuss the lessons learned over eight years of teaching the corresponding courses and compare the results of our student project to the trends summarised in the recently published approached from other universities and countries. \\
~\\
\emph{Preprint. Accepted to the 20th International Conference on Mobile and Ubiquitous Systems: Computing, Networking and Services (MobiQuitous 2023), Melbourne, Australia, November, 2023. Springer. Final version to be published by Springer (In Press).}

\end{abstract}
\section{Introduction}

To develop sustainable application that fulfils users' needs, it is crucial to understand the whole range of user experiences and expectations  \cite{anderson2016mobile,hoehle2015mobile}.  
The developers require to understand users' motivations and what they perceive as useful and easy to use \cite{amin2014user,kim2013study}, but this is a complex task which is not easy to master. Even when a developer follows general guidelines and best practices, e.g., the {human interface guidelines} specified by Apple\footnote{\url{https://developer.apple.com/design/human-interface-guidelines}} or the design guidelines specified by Android\footnote{\url{https://developer.android.com/design/ui/mobile/guides/foundations/accessibility}}, the needs of target user group should be analysed in addition.  
One of the approaches to understand needs of particular group of users, is to use so-called ``personas'' presenting composite user archetypes that covers  consist of users’ ways of thinking, behaviour, goals, and motivations for a specific context~\cite{cooper2014face,liu2022curated}. The latest systematic mapping study on the use of personas in Requirements engineering was introduced in \cite{karolita2023use}.

Having theoretical knowledge on the importance of these aspects isn't as effective as learning from real-life scenarios. Real-world problems serve as the catalyst and central point for student engagement, which was confirmed by numerous studies from many areas of software and system development, see for example \cite{boud2013challenge,savery2015overview,Armarego2002}. Collaborative active learning allows students to explore how the theories learned in the class apply in real-world context
\cite{dharmaratne2022implementing}. 
This has an especially strong effect when the real-life scenarios aren't simulated, i.e. when the work is conducted with real stakeholders for the currently existing problems that require solutions as soon as possible.   

\textbf{Contributions:}  We present a discussion on project-based capstone course provided at the RMIT University   in the School of Computing Technologies with the focus on mobile applications and mobile user experience. 
We introduce how it is embedded into the curriculum, as well as discuss lessons learned over eight years of teaching the corresponding courses and discuss two examples of students' projects, which have been conducted in the area of mobile app development.

\textbf{Outline:} The rest of the paper is organised as follows. 
Section~\ref{sec:related} presents an overview of related work. 
Section~\ref{sec:projects} presents a number of examples of the student projects as well as a discussion on our observations and lessons learned, comparing them with results of other studies. 
Finally, Section~\ref{sec:conclusions} summarises the paper. 

\section{Related work}
\label{sec:related}

A systematic review of teaching methods in software engineering was presented in \cite{anivcic2022teaching}. In that study, the authors summarised the most common teaching methods with the aim of a better alignment of Software Engineering (SE) courses with industry-relevant knowledge, tools, and practices.

Project-based learning offers numerous benefits to students. When the study projects are either derived from the industry or mirror genuine industrial challenges, the learning tasks naturally adopt a problem-based approach. 
By the same reasons, many universities nowadays transform the teaching activities to be \emph{studio-based}, i.e., to focus on learning through hands-on activities and developing as result particular artefact, such as UI mock-ups, Web or mobile app prototypes, project management artefacts, etc. The core difference between studio-based courses and capstone courses is in the level of the study: studio-based leaning in general doesn't have any limitations on the the year of study and can be used for introductory courses as well. 

In contrast to this, capstone courses are the final year (final semester) courses, which aim is  to integrate the whole set of skills and knowledge obtained by the students within the previous semesters, and apply this integrated set within a (relatively) large project.
Also, these projects are especially appropriate for so-called Work-Integrated Learning (WIL), where an opportunity of obtaining an industry experience is provided to the students ~\cite{ICSE_2018_capstone,balaban2018software,knudson2018global,bastarrica2017can}. %
 Another notable benefit  is offering experience of collaboration within a team, helping them develop skills in coordinating and sharing responsibilities within a group~\cite{schoor2015regulation,gol2007collaborative,panitz1999collaborative}.

There are also approaches that aim to introduce students to research in computer science to retain them in academia or to prepare them to further study~\cite{peckham2007increasing,cooper2010teaching}. 
We propose to provide an optional extension to the course, where the focus will be on research activities related to the  project the students worked on within the semester. 
These small research projects could be very short (up to two weeks, to be completed after the teaching / assessment period of the semester) and preferably  sponsored by industrial partners. Several examples of such projects have been presented in \cite{christianto2018enhancing,clunne2017modelling,gaikwad2019voice,george2020usage,spichkova2020vm2,sun2018software,spichkova2019comparison,spichkova2020gosecure,spichkova2020ICSoft,ICECCS_SMI,spichkova2019easy}.
 
In  our previous work \cite{simic2016enhancing,spichkova2017autonomous,young2021project}, 
we investigated solutions to include in the STEM (Science, Technology, Engineering and Mathematics) Bachelor and Master programs a number of  research activities within 
eHealth, bio-engineering and autonomous systems.  
In our other work \cite{spichkova2019industry},  we summarised the  lessons learned as well as ideas to
redesign a capstone course, where we had to deal with many diverse challenges (e.g.,  large cohorts).
In our current work, we focus on projects related mobile user experience and aim to compare the results of our student project with the trends summarised in the recently published approached from other universities and countries.

\section{Project-based learning: mobile applications}
\label{sec:projects}  

The project-based capstone courses 
have been conducted in our case within Bachelor of SE (BP096) and Master of Information Technology (MC208) study programs. 
The RMIT academics provided their expertise on PM and SE, especially focusing on requirements engineering (RE) aspects, analysis of the provided problem and its scope, communication of the clients, etc. where some further domain expertise was also provided by our industrial partners. 
We typically allocated five to six students to each team (if required by the project scope, in few cases we allowed  teams of four or seven students). 

As the pre-requisite to this capstone course, each student had to pass the courses on fundamentals of SE, programming and basic PM.  
In that capstone course the students apply their Agile/ Scrum (as well as general PM) skills as well as their previously acquired knowledge on software development methodologies in real-life settings, assuming that the core concepts have been already studied in the prerequisite courses.   
The structure of this project-based course was evolving over the last eight years, to incorporate the industry shift to Agile/Scrum development as well as to accommodate the recent needs for remote/hybrid working environment. 

Generally, the overall set of project topics that we provide within the course isn't limited to the projects on mobile app development: we provide a wide range of topics to cover several application areas as we assume that this will help to  increase students' engagement. However, mobile app development and related projects build a large proportion of the proposed capstones. 
In this paper,  we selected two examples of the projects that focused on mobile app development.
  
\subsection{Deck Logger App}

This project was conducted in Melbourne, Victoria (Australia) in collaboration with Spatial Vision. The aim of the project was to develop a prototype of a mobile app for Australian fishers. 
The mobile user needs identified in this project are largely different from findings presented in \cite{kanij2023developing} by Kanij et al. based on a case study of fisherfolk communities in Bangladesh. These differences highlight the influence of human aspects and national culture (see also~\cite{alsanoosy2019cultural,alsanoosy2019detailed,alsanoosy2020does,alsanoosy2020identification,grundy2020towards,grundy2020humanise,LIU2023111791,shahin2022operationalizing,shams2020society}), and the impacts of low socio-economic background of users. 

\begin{figure}
\begin{center}
\includegraphics[width=0.25\textwidth]{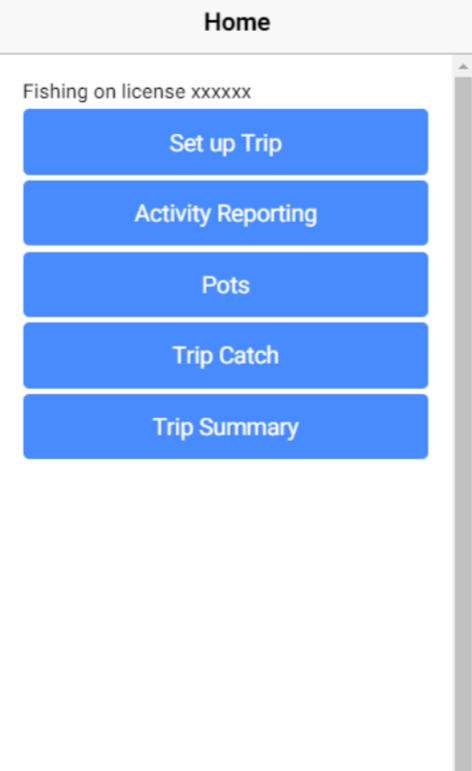}~~~~~~
\includegraphics[width=0.25\textwidth]{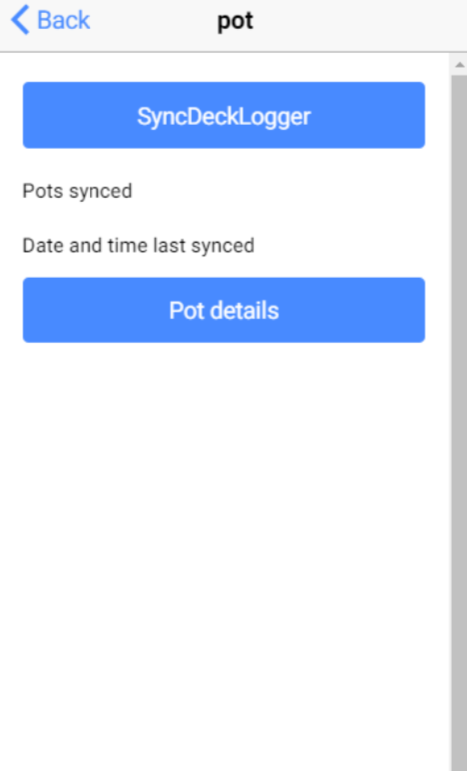}~~~~~~
\includegraphics[width=0.25\textwidth]{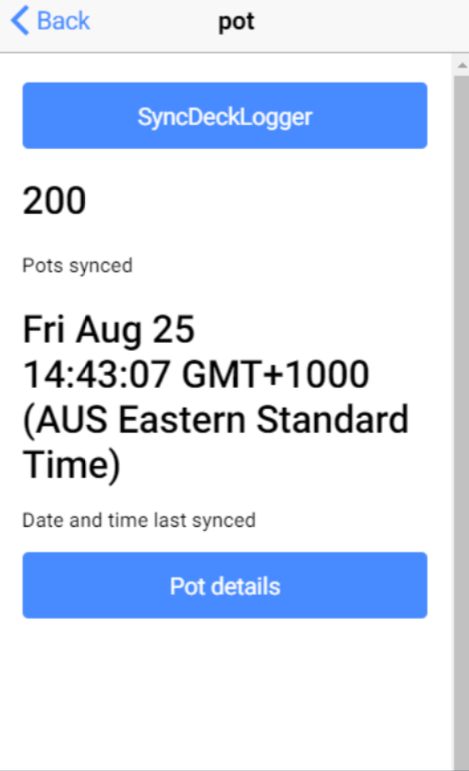}
\end{center}
\caption{Samples of the app screens for the Deck Logger App prototype: Home Screen After login, pot details page before syncing with deck logger, and pot details page after syncing with deck logger} 
\label{fig:screens1}
\end{figure}

Some of the usability aspects critical for mobile user experience have been identified in both studies, such as simple (or even simplistic) user interface with large buttons, see Figure~\ref{fig:screens1}. 
However, these aspects are typical for apps that should be used in environments where a user has limited time to deal with application and might have not very stable (physical) standing position, like on a boat, construction side, etc.
However, in our study  we haven't identified any specific requirements related to literacy, digital literacy, lack of access to smartphones, etc. For example, one of the complexities with using mobile apps identified by by Kanij et al. was that most of Bangladeshi fishermen could not read the text, which made the feature ``text to voice'' very important in this context.
In contrast to this, Australian fishers didn't have any issues with literacy and digital literacy, have smartphones, and therefore required almost completely different set of functionalities: an application to automate reporting process.

Fisheries Victoria requires Rock Lobster fishers to report catch-and-effort data for each fishing trip which they undertake. For each trip, fishers must submit three types of reports: a pre-fishing report, a pre-landing report, and post-landing report. When these reports are manually completed by filling out an online or paper-based logs depending on the type of report, this becomes a labour intensive and time-consuming activity for fishermen.

At the time of the project, the Victorian Rock Lobster industry applied electronic handheld devices produced by New Zealand based technology company Zebra-Tech to record catch-and-effort data. These handheld devices (so-called ``Deck Loggers''\footnote{\url{https://www.zebra-tech.co.nz/deck-logger/}} are easy-to-use data collection tools that require manual entry of data. However, these devices didn't offer any reporting functionality and thus manual extraction and storage of the data is required. 
Zebra-Tech also produces a scientific data collection device called a ``Wet Tag'' that can be attached various pieces of fishing equipment, e.g., a Rock Lobster Pot. These devices record location, temperature, depth and soak-time, which can be automatically transferred via Bluetooth from the Wet Tag to the Deck Logger.

The high-level objectives of this capstone project were to create a working prototype of a mobile application which:
\begin{itemize} 
\item Enables retrieval and processing of data from existing Zebra-Tech hardware,
\item Automates/streamlines the current mandatory reporting process, to decrease the time and effort fishers spend completing the reporting process,
\item Enables map-based and statistical summaries of data.
\end{itemize}
different to the objectives discussed by Kanij et al. \cite{kanij2023developing} in context of the case study of fisherfolk communities in Bangladesh. The reason for this was a large difference in the needs of fishers in Australia and Bangladesh, which is based on the differences in their technical and social background and industrial policies.

The app was developed using Ionic, a hybrid mobile technology that enables the application to be built for both Android and IOS using one code base. Ionic 3 was selected over other available hybrid frameworks as team members were familiar with the associated technologies and languages used within the Ionic framework. Ionic provides native user interface components for allowing for rapid prototyping and development.

\begin{figure}[t!]
\begin{center}
\includegraphics[width=0.9\textwidth]{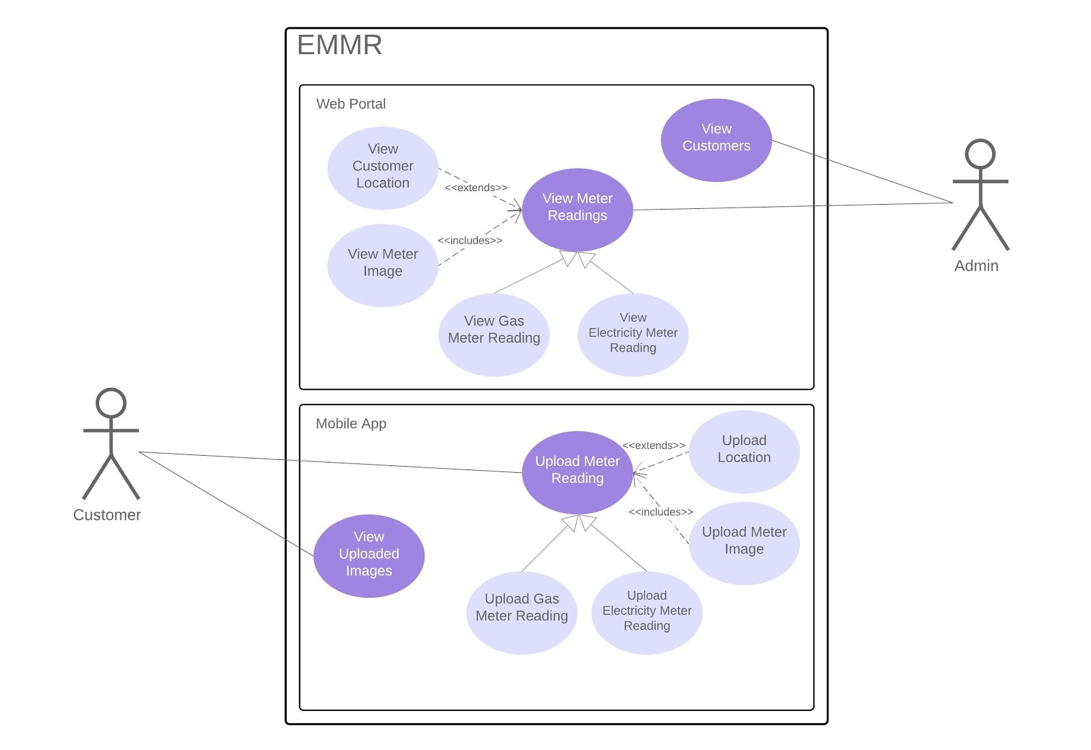} 
\end{center}
\caption{Mobile Meter Reading: Use case diagram}
\label{fig:EMMR-us}
\end{figure}
  
\subsection{Mobile Meter Reading for Non-Smart Meters}

The project was conducted in collaboration with Shine Solutions Group Pty Ltd and Energy Australia, which is an electricity and gas retailing private company in Australia. It supplies electricity and natural gas to more than 2.6 million residential and business customers throughout Australia. 
This project had two focal points: 
\begin{itemize}
    \item 
    Mobile app development conducted by students within the capstone course, which we discuss in this paper, and  
    \item 
    Application of computer vision approaches (which was mainly conducted within an additional research project). The results of the corresponding research project, which was focused on analysis of several computer vision approaches to allow efficient automation, have been presented in \cite{spichkova2020ICSoft,spichkova2019easy}.
\end{itemize}
 In this paper we focus on the aspects related to the mobile user experience as well as on the corresponding learning and teaching aspects.

At the time of the project, consumers used to update  their utility reading  manually, using the corresponding website.  
Updating the data fully manually might be complicated and time-consuming for the consumers as they need to enter a lot of data. Also, the meters are often dusted, which makes it not so easy to read the numbers, especially for elderly people or for people with vision impairment. The cases when some digits are clipped are especially complicated. 

The students' task was to elaborate 
a simple automated accessible  alternative, so that the consumers can use a simple app on their personal smartphone for capturing the meter readings in a semi-automated way.  The proposed scenario was simple: a consumers opens the app, uses smartphone's camera to obtain an image of the meter, and the app will automatically identify from this image  the current readings of the meter, where all unrelated data will be filtered out. Then the best suited results are presented on the mobile app's screen. If the consumer accepts the identified readings as acceptable, they submit them to the Energy Australia system by clicking the corresponding button in the app. Otherwise, they might repeat an attempt.

\begin{figure}
\begin{center}
\includegraphics[width=0.28\textwidth]{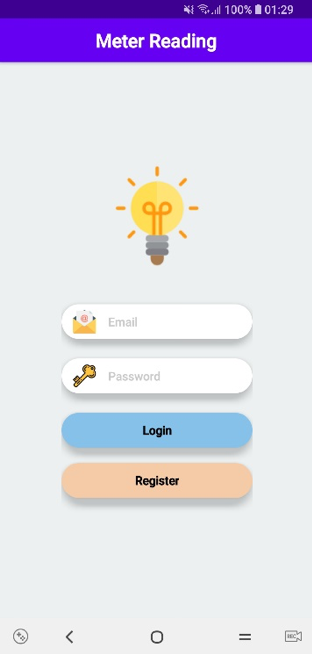}~~~
\includegraphics[width=0.285\textwidth]{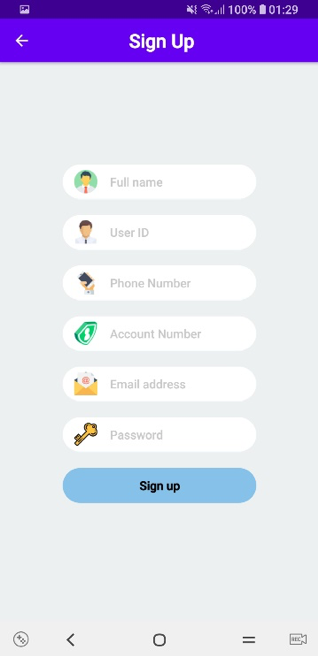}~~~
\includegraphics[width=0.30\textwidth]{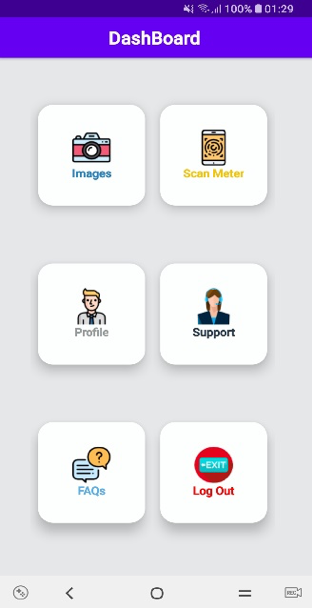}
\\~~~\\
 
\includegraphics[width=0.28\textwidth]{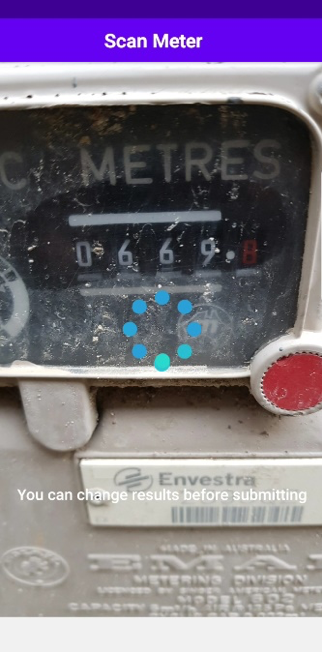}~~~ 
\includegraphics[width=0.29\textwidth]{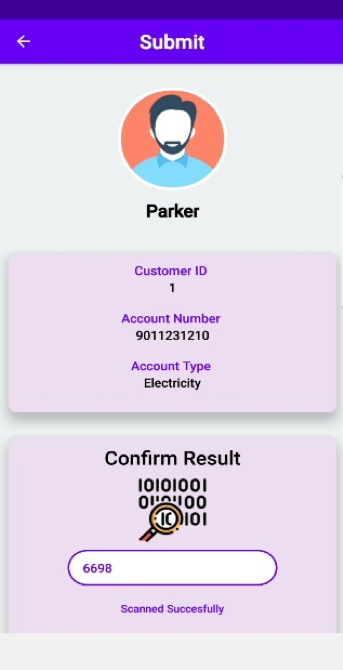} ~~~
\includegraphics[width=0.30\textwidth]{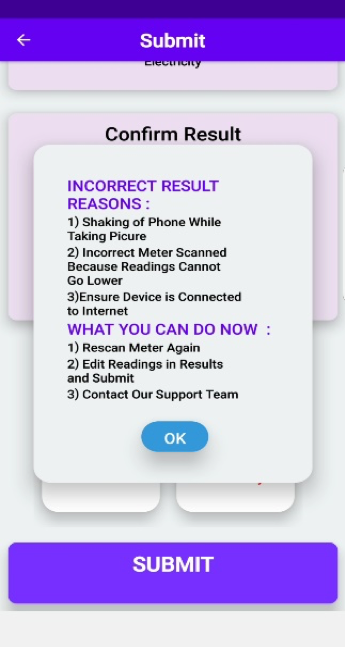}
\end{center}
\caption{Samples of the app screens: Login page, Sign Up page, User Dashboard, Capturing of the meter reading, Presenting correct results, Providing hints in the case of incorrectly identified readings}
\label{fig:EMMRscreens2}
\end{figure}

Figure~\ref{fig:EMMR-us} presents a use case diagram for the app.  
Figure~
\ref{fig:EMMRscreens2} presents examples of some pages of the app. 
The app prototype was developed as an Android application built using React Native.

\section{Discussion and Conclusions}
\label{sec:conclusions}

There are many studies on project-based learning of software development. However, we believe that there are many unexplored / not discussed aspects that need to be analysed to improve learning and teaching process. In the areas like mobile user experience, where the perception of quality and the corresponding expectations are constantly evolving over recent years, the discussion of latest lessons learned is especially important. 
An experience report on using project-based learning in a mobile application development course was presented in \cite{francese2015using}. The core lessons learned, identified by authors, aren't specific for mobile app development, and are rather related to general project management aspects. Similar observations have been presented in
\cite{lopez2023applying}, where a a case study on applying user experience and user-centred design approaches in teaching of mobile app development  was introduced.  
We agree that these points are critical,  especially when the project are conducted using Agile/Scrum. 
Our observations also confirm that many students expect a software development process to be a waterfall process with no changes to the requirements specification over the project duration. 
It is only once they engage in an actual project with a real industrial partner that they come to acknowledge the possibility of evolution/refinement of many elements throughout the course of the project: from functional and non-functional requirements to system architecture or user interface.  

It is generally hard for students (and potentially for for novice software engineers in general) to comprehend the above points, as typically the assessment tasks in school and in the first year course on university level are  well-formulated and well-structured from the beginning. In real life this is almost never the case, as requirements should be elicited and confirmed, but then can evolve over the duration of the project. This is especially critical for the requirements related to user experience, as in many cases stakeholders might be able to provide detailed feedback only after engaging with an early prototype of the solution.
Even when 
\begin{itemize}
    \item 
the importance of these points is discussed in detail in RE literature, see e.g.,  \cite{nuseibeh2000requirements,sommerville1997requirements}, and 
\item 
the students are taught the corresponding material within the software engineering courses, 
\end{itemize}
as soon it comes to a real project with real stakeholders and real user experience issues, many students are deeply surprised that the adjustments are needed, that the interface might need to be re-done many times, that additional functional and especially non-functional requirements are specified as top-priority in the middle or even towards the end of the project. 

Thus, our observations demonstrate that the best way to master 
these requirements engineering and user experience aspects is to conduct a real project within a capstone course. However, it is also important to mention that completing pre-requisite courses strictly before doing the capstone is crucial, because supervisors can only mentor the students on this stage and if the students missed a large portion on background knowledge it might be to hard to prepare.  

As mentioned in~\cite{bruegge2015software},  conducting these project with real industrial clients provides better  learning effect. 
Our observations confirmed the points discussed in~\cite{rodriguez2016measuring} on enriching the students' software development project with agile coaching. Based on our lessons learned, one of the successful strategies is to enforce having an adjusted Scrum routine for meetings: 
\begin{itemize}
    \item It is important to remind students on having daily stand-ups and to update regularly their project board (Trello or Jira, depending on what the team / industrial partner decided to curate).
    \item If a team struggles with presentations and explanations in the Sprint review meeting with industrial partners (where the current results should be demonstrated), an additional internal (without the industrial partners) meeting with mentor might be useful, e.g. 0.5-1h before the main meeting. 
    \item An internal Sprint retrospective meeting after the Sprint review meeting is a critical must-have part. It might be as short as 0.5h (or even shorter if the team performs really well) but having a regular catch up and regular opportunity to discuss points to improve is very effective for students to be on track and and also to share their feelings and concerns with the mentor.
\end{itemize}
Thus, from what we observed over the last eight years of supervision / coordination of these projects, it's also critical to provide within the course regular mentoring activities to the students on the aspects they struggle most.

\section*{{Acknowledgements}}
We would like to thank  Shine Solutions Group Pty Ltd and Spatial Vision for supporting the projects.
We also would like to thank students who participated in the discussed projects: 
Callum Pearse, 
Baoyun Chen,
Shiran Ekanayake, 
Lijith Win Vijay,
Harish Narayanasamy,
Ashish Bhardwaj,  
Johan van Zyl,  
Lavanya Krishnamurthy, 
Nirav Desai, and
Siddharth Sachdeva.

\bibliographystyle{splncs04}
%

\end{document}